%Paper: hep-th/9502162
%From: dhoker@physics.ucla.edu (D'Hoker)
%Date: Tue, 28 Feb 1995 10:28:40 +0800

%
%%%%%%%%%%%%%%%%%%%%%%%%%%%%%%%%%%%%%%%%%%%%%%%%%%%%%%%%%%%%%%%%%%%%%%%
%%%%%%%%%%%%%%%%%%%%%%%%%%%%%%%%%%%%%%%%%%%%%%%%%%%%%%%%%%%%%%%%%%%%%%%
%%%%%%%%                                                         %%%%%%
%%%%%%%%  Invariant Effective Actions,                           %%%%%%
%%%%%%%%  Cohomology of homogeneous Spaces and Anomalies         %%%%%%
%%%%%%%%                                                         %%%%%%
%%%%%%%%  by                                                     %%%%%%
%%%%%%%%                                                         %%%%%%
%%%%%%%%  Eric D'Hoker                                           %%%%%%
%%%%%%%%                                                         %%%%%%
%%%%%%%%                                                         %%%%%%
%%%%%%%%  Typeset in Plain TeX, no macros needed                 %%%%%%
%%%%%%%%                                                         %%%%%%
%%%%%%%%                                                         %%%%%%
%%%%%%%%%%%%%%%%%%%%%%%%%%%%%%%%%%%%%%%%%%%%%%%%%%%%%%%%%%%%%%%%%%%%%%%
%%%%%%%%%%%%%%%%%%%%%%%%%%%%%%%%%%%%%%%%%%%%%%%%%%%%%%%%%%%%%%%%%%%%%%%

\magnification=\magstep1
\baselineskip=14pt
\overfullrule=0pt

\def\no{\noindent}

\def\G{{\cal G}}
\def\H{{\cal H}}
\def\M{{\cal M}}
\def\half{{1 \over 2}}
\def\O{{\cal O}}
\def\tr{{\rm tr}}

\rightline{UCLA/95/TEP/5}

\bigskip
\bigskip

\centerline{{\bf INVARIANT EFFECTIVE ACTIONS,}}
\medskip
\centerline{{\bf COHOMOLOGY OF HOMOGENEOUS SPACES}}
\medskip
\centerline{{\bf AND ANOMALIES}}

\bigskip
\bigskip
\bigskip
\bigskip

\centerline{{\bf Eric D'Hoker}
\footnote{$^*$} {E-mail address : dhoker@uclahep.physics.ucla.edu;
 Research is supported in part by National Science Foundation grant
PHY-92-18990.}}

\medskip

\centerline{{\it Physics Department}}
\centerline{{\it University of California Los Angeles}}
\centerline{{\it Los Angeles, CA 90024, USA}}

\bigskip
\bigskip
\bigskip

\centerline{{\bf Abstract}}

\bigskip
\bigskip

We construct the most general local effective actions for Goldstone boson
fields associated with spontaneous symmetry breakdown from a group $G$ to a
subgroup $H$. In a preceding paper, it was shown that any $G$-invariant
term in the action, which results from a non-invariant Lagrangian density,
corresponds to a non-trivial generator of the de Rham cohomology classes of
$G/H$.  Here, we present an explicit construction of all the generators of this
cohomology for any coset space $G/H$ and compact, connected group $G$.
Generators contributing to actions in 4-dimensional space-time arise either as
products of generators of lower degree such as the Goldstone-Wilczek current,
or are of the Wess-Zumino-Witten type. The latter arise if and only if $G$
has a non-zero  $G$-invariant symmetric $d$-symbol, which vanishes when
restricted to the subgroup $H$, i.e. when $G$ has anomalous representations
in which $H$ is embedded in an anomaly free way.
 Coupling of additional gauge fields leads to actions whose gauge variation
coincides with the chiral anomaly, which is carried here by Goldstone
boson fields at tree level. Generators contributing to actions in
3-dimensional space-time arise as Chern-Simons terms evaluated on connections
that are composites of the Goldstone field.

\vfill\eject

\baselineskip=17pt

\no
{\bf 1. Introduction}

\medskip

Local effective field theories are widely used to capture the dynamics of
Goldstone bosons resulting from the spontaneous breakdown of continuous
symmetries. The power of local effective field theories lies in the fact that
their structure is largely determined by symmetry considerations alone.
If the symmetry of the action is a (compact) group $G$,
and the symmetry of the vacuum is a subgroup $H$, then the Goldstone
fields  $\pi ^a (x)$ parametrize the coset space
$G/H$ with $a=1, \cdots, {\rm dim}~ G/H$.
The corresponding effective field theory depends on only a finite number of
couplings, up to any given order in an expansion in powers of derivatives.

General methods for constructing invariant actions were given in [1] for
$SU(2)_L \times SU(2)_R$ and extended to the case of arbitrary $G$ and $H$
in [2]. These methods are based on producing the most general
invariant Lagrangian density, but do not consider exceptions where the
action is invariant under $G$ while the Lagrangian density transforms with a
total derivative term.
The Wess-Zumino-Witten (WZW) term, which was originally considered as an
effective action for chiral anomalies, is an example of such an exception
[3,4].

The success of the effective action approach relies upon the  assumption
that we can enumerate all possible invariant contributions to the action.
In a recent paper [5], we addressed the issue of constructing all invariant
contributions to the effective action for arbitrary $G$ and $H$.
We exhibited a one to one correspondence
between local terms in the effective Lagrangian density, that although not
$G$-invariant, yield $G$-invariant contributions to the effective action, and
generators of the de Rham cohomology classes of the coset space $G/H$.
The invariant effective action $S[\pi]$ in space-time dimension $n-1$ is given
in terms of a de Rham cohomology generator $\Omega $ of degree $n$  as
follows
$$
S[\pi] = \int _{B_n} \Omega (\tilde \pi)
\eqno (1.1)
$$
Here, $n-1$ dimensional space-time $M_{n-1}$ is extended to an $n$-ball $B_n$
with boundary $M_{n-1}$, and the field $\tilde \pi$ interpolates between the
original field $\pi$ on $M_{n-1}$ and the field whose value is zero at all
space-time points. Each independent cohomology generator
produces a $G$-invariant action given by (1.1) which arises
from a non-invariant Lagrangian density
and escapes the construction of $G$-invariant actions given in [2].
In four dimensions, we are interested in de Rham cohomology
generators of degree 5.
For the simplest case of $G/H = SU(N)$, $N\geq 3$ for example,
there is just a
single generator which corresponds to the original WZW term.

In the present paper, we derive the general structure of the de
Rham cohomology classes for arbitary compact, connected $G$ and
arbitrary subgroup $H$. We emphazise the classes of degrees 4 and
5, which are the ones that yield invariant actions in 3- and 4-dimensional
space-times respectively.  We exhibit a correspondence between cohomology
generators and invariant symmetric tensors, which is closely related to
the transgression map introduced by A.~Weil, H.~Cartan and A.~Borel  [6,7]
and to the descent equations that arose
in the study of chiral anomalies [8,9,10].
More specifically, when a cohomology generator on $G/H$ contains
linear combinations of
products of cohomology generators on $G/H$ of lower degree, the corresponding
generator is said to be {\it decomposable}; in the contrary case, the generator
is said to be {\it primitive}. The full cohomology of degree $n$ can be
obtained simply from the primitive cohomology generators of all degrees up
to and including $n$. For the case of most interest, cohomology of degree 5,
we need the primitive generators of degree 5, and also all primitive
generators of degrees 1, 2, 3 and 4, which we explicitly construct.
We show that primitive cohomology generators are in one to one correspondence
with $G$-invariant rank 3 symmetric tensors which vanish when restricted to
the generators of the subalgebra of $H$.  \footnote{*} {This general result is
the one obtained in [7,11].}  The primitive cohomology generators then always
arise from non-trivial cohomology generators on $G$ that can be consistently
projected down to $G/H$. While the analysis of cohomology
classes of homogeneous spaces has been the subject of intense study in
mathematics, the number and the form of the generators does not seem to be
available in explicit form for general $G/H$.

The elements in the above construction have direct physical counterparts.
Cohomology generators of degree 2 arise in connection with the dynamics of
electrically charged particles moving in the presence of a magnetic
monopole field. (See e.g. ref. [4])
Cohomology generators of degree 3 are associated with the
Goldstone-Wilczek current which plays an important role in the study of
quantum numbers carried by solitons [12]. The same generators are also
familiar as the WZW term in 2 space-time dimensions.
Cohomology generators of degree 4 are given by the second Chern class,
evaluated on a composite $\H$-valued gauge field on $G/H$. The resulting
invariant action may be recast directly in three dimensions and turns out to
be given by the Chern-Simons form [8,13] evaluated on the same
$\H$-valued gauge field. This construction of 3-dimensional invariant
actions generalizes the Hopf type invariants.

The primitive cohomology generators of degree 5 are related to the WZW term.
The $G$-invariant symmetric tensors of rank 3 are identical to the $d$-symbols
encountered in the study of chiral anomalies in quantum field theories with
chiral fermions coupled to non-Abelian gauge fields [13,14]. Thus,
primitive cohomology generators of degree 5 occur when the group $G$ has at
least some representations in which the $d$-symbol is non-vanishing; in this
case the group $G$ by itself admits a WZW term.  The corresponding $d$-symbol
must vanish upon restriction to the Lie subalgebra of $H$, so that the WZW
term  for $G$ can be gauged with respect to the anomaly free subgroup $H$, and
projected down to a well-defined action where the field takes values in $G/H$.
The fact that the WZW term can be constructed this way has been known for
some time [10,15]. What we show here is that this construction produces {\it
all Lagrangian densities} that, although not $G$-invariant, yield
$G$-invariant actions.

As is well-known, WZW terms cannot in general be coupled to gauge fields
in a locally $G$-invariant way.
This failure to maintain gauge invariance in a theory with chiral fermions is
the chiral anomaly, and the WZW term was introduced precisely as an
effective action reproducing this anomaly [3].
It is remarkable that we have here identified a source of anomalies in a
completely classical theory with Goldstone bosons only, without ever mentioning
chiral fermions or short distance divergences that underly chiral anomalies.
This result seems to emphasise again the fundamental connection between chiral
anomalies and spontaneous symmmetry breakdown [16].

\medskip

The remainder of this paper is organized as follows. In \S 2, we review
standard results on the differential calculus of homogeneous
spaces and outline the construction of cohomology generators carried out
in the paper.
In \S 3, we reduce the cohomology problem to the
classification of constant symmetric tensors on $G$ with certain invariance
and vanishing properties with respect to $G$ and $H$.  In \S 4, we integrate
the equations for the cohomology of degrees 2, 3 and 4 with simply
connected $G/H$. In \S 5, we  analyze the cohomology of degree 5 for simply
connected $G/H$, and relate  primitive cohomology generators of $H^5(G/H;R)$ to
$G$-invariant symmetric tensor of rank 3 on $G$, that vanish upon restriction
to
the Lie algebra of $H$. We show that generators either factorize into
generators of lower degree or are obtained from the projection onto $G/H$ of
primitive generators of $G$. In \S 6, we extend our results to include the
case of non-simply connected $G/H$. Finally, in \S 7, we discuss gauging the
$G$-invariant actions constructed here.

\bigskip
\bigskip

\no
{\bf 2. Differential Calculus on Homogeneous Spaces G/H}

\medskip

We assume that $G$ is a compact connected Lie group with $H$ a subgroup and we
denote the corresponding Lie algebras by $\G$ and $\H$ respectively.
On a homogeneous space, the Lie algebra $\G$ may be decomposed as
follows $\G = \H + \M$, with
$$
[\H, \H ] \subset \H, \qquad [\H, \M ] \subset \M
\eqno (2.1)
$$
We shall use a notation in which the indices for the generators of $\G$ are
Latin capital letters $A=1, \cdots ,\dim G$ and are decomposed into a set of
Latin lower case letters $a$ running over the generators of $\M$ and a set
of Greek lower case letters $\alpha $ running over the generators of $\H$.

It is a fundamental result
\footnote{*} {Introductions to differential geometry and
cohomology of Lie groups and homogeneous spaces may be found in [11,17] and
especially in [18]. Following standard physics notation, wedge products will
be understood for the multiplication of differential forms.}
 that the cohomology of homogeneous spaces $G/H$
is given by the classes of closed $G$-invariant differential forms on $G/H$
modulo forms that are the exterior derivative of $G$-invariant forms on $G/H$.
To construct $G$-invariant differential forms, we begin by introducing
a basis of left-invariant differential forms of degree 1 on $G$, denoted by
$\theta ^A,\ A=1,\cdots,\dim G$ and expressed in terms of the Goldstone field
$U$ as follows :
$$
\theta = \theta ^A T^A = U^{-1}dU, \qquad\qquad [T^A,T^B]=f^{ABC}T^C
\eqno (2.2)
$$
Here $T^A$ are matrices in the representation under which $U$ transforms and
$f$ are the totally anti-symmetric structure constants of $G$.
Throughout, repeated indices are to be summed over and we shall not
distinguish between upper and lower indices.
{}From the definition of $\theta$, we see that its components satisfy the
Maurer-Cartan equations
$$
d\theta ^C + \half f^{ABC} \theta ^A \theta ^B =0
\eqno (2.3)
$$
We may use the basis of left-invariant differentials $\theta ^A$ to exhibit
general differential forms of degree $n$ on $G$.
$$
\Omega = {1 \over n!} \omega _{A_1 \cdots A_n} \theta ^{A_1} \cdots
\theta ^{A_n}
\eqno (2.4)
$$
For arbitrary differential forms $\Omega$, the coefficients
$\omega _{A_1 \cdots A_n}$ are arbitrary functions of $G$.
However, if we restrict to $G$-invariant differential forms that are
well-defined on $G/H$, we have the following Theorem [17] which puts strong
conditions on the coefficients $\omega _{A_1 \cdots A_n}$.

\medskip

\no
{\bf Theorem 1}

\medskip

{\it The ring of invariant forms $\Omega$ on $G/H$ is given by the exterior
  algebra of multilinear anti-symmetric maps on $\G$ which vanish on $\H$
		and which are invariant under the adjoint action of $\H$.}

\medskip

This means that the coefficients $\omega _{A_1 \cdots A_n}$
must be constant as functions on $G$, vanish whenever one of the indices $A_i$
corresponds to a generator in $\H$ and be invariant under the adjoint action
of the group $H$ (or $H$-invariant for short). As a result, the ranges of the
summation indices in (2.4) may be restricted to $\M$ and we have
$$
\Omega = {1 \over n!} \omega _{a_1 \cdots a_n} \theta ^{a_1} \cdots
\theta ^{a_n} \qquad \qquad \omega _{c[a_2 \cdots a_n} f_{a_1]} {}^{c\beta }
=0 \eqno (2.5)
$$
Here the symbol $[\ ]$ denotes anti-symmetrization of the indices inside the
brackets and the index $\beta $ corresponds to any generator of $\H$.

This theorem is easily understood in view of the  $G$-transformation
properties  of the basic differential forms of (2.2).  Under left $G$
transformations, the Goldstone fields transform as follows [2]: $U(\pi) \to
gU (\pi) = U(\pi') h(g,\pi)$ where $h(g,\pi) \in H$.  As a result, the forms
$\theta ^a$ transform in the adjoint representation of $\H$, whereas $\theta
^\alpha$ transform as a $\H$-valued connection :
$$
\eqalign{ \theta ^a T^a  &\mapsto h^{-1} \theta ^a T^a h \cr
\theta ^\alpha T^\alpha & \mapsto h^{-1} \theta ^\alpha T^\alpha h
								+h^{-1}dh \cr}
\eqno (2.6)
$$
Thus, $G$ transformations on
$U$ induce $\H$-valued gauge transformations on $\theta$, and a $G$-invariant
form on $G/H$ should be invariant under $\H$-gauge transformations; this is
precisely what Theorem 1 guarantees.

We also need differential forms of degree $n$ which are tensors of
rank $m$  on the Lie algebra $\G$ (called tensor forms), defined as follows
$$
\Omega _{B_1\cdots B_m} = {1 \over n!} \omega _{B_1\cdots B_m;A_1\cdots A_n}
\theta ^{A_1} \cdots \theta ^{A_n}
\eqno (2.7)
$$
Such a tensor form
is $G$-covariant and well-defined on $G/H$ if the coefficients
$\omega _{B_1\cdots B_m;A_1\cdots A_n}$ are constant, vanish whenever one
of the $A_i$ indices corresponds to a generator of $\H$, and -- when viewed as
a tensor of rank $m+n$ -- are invariant under the action of $\H$ in the adjoint
representation. We shall not at present assume any particular symmetry amongst
$B_i$ indices or between $B_i$ indices and $A_j$ indices.

An operation $\O$ on differential forms $\Omega _1$ and $\Omega _2$ of degrees
$d_1$ and $d_2$ respectively, satisfying the following rule
$$
\O (\Omega _1  \Omega _2) = (\O \Omega _1)  \Omega _2 +
(-)^{w_1} \Omega _1  (\O \Omega _2)
\eqno (2.8)
$$
is said to act as a {\it derivative} if $w_1=0$ and as an
{\it anti-derivative} if $w_1=d_1$.

The exterior derivative $d$ on differential forms with constant coefficients is
defined by the {\it anti-derivative rule} in (2.8) and the
Maurer-Cartan equations  of (2.3). For example on tensor forms of (2.7) we have
$$
d\Omega _{B_1 \cdots B_m} = -{1 \over 2(n-1)!}\ \omega _{B_1 \cdots B_m; A_1
\cdots A_n}\ f^{A_1BC}\ \theta ^B \theta ^C \theta ^{A_2} \cdots \theta ^{A_n}
\eqno (2.9a)
$$
On $G$-invariant forms on $G/H$, as defined by Theorem 1 and (2.5), the action
of the exterior derivative takes on a simplified form :
$$
 d\Omega = -{1 \over 2(n-1)!}\ \omega _{a_1 \cdots a_n}\
f^{a_1bc}\ \theta ^b \theta ^c \theta ^{a_2} \cdots \theta ^{a_n}
\eqno (2.9b)
$$
On tensor forms of rank $m$, it is natural to act with the covariant
derivative $D$, defined by the {\it anti-derivative rule} of (2.8) and its
expression on tensor forms of rank 1. On tensor forms of rank $m$ we have
$$
D\Omega _{B_1 \cdots B_m} = d\Omega _{B_1 \cdots B_m} +
f_{B_1 BC} \theta _B \Omega _{C B_2 \cdots B_m} + \cdots +
f_{B_m BC} \theta _B \Omega _{ B_1 \cdots B_{m-1}C}
\eqno (2.10)
$$
Since the connection $\theta $ has zero curvature, we have
$D^2=0$ on tensor forms of all ranks. The exterior derivative
increases the degree by one unit and leaves the rank of the tensor form
unchanged.

We shall also make use of the fundamental operation $i_A$,
which lowers the degree by one and increases the rank by one :
$$
i_A \Omega _{B_1 \cdots B_m} = {1 \over (n-1)!} \omega _{B_1
\cdots B_m;A A_2 \cdots A_n}
	\theta ^{A_2} \cdots \theta ^{A_n}
\eqno (2.11)
$$
It is easy to see that $i_A$ acts with the {\it anti-derivative rule} of (2.8)
and satisfies $i_A i_B + i_B i_A =0$ on all forms.

Rotation under the action of $\G$ obeys the {\it derivative rule} of (2.8)
and acts on tensor forms of rank $m$ by
$$
L_A \Omega _{B_1 \cdots B_m}  = f_{AB_1B}\ \Omega _{B B_2 \cdots B_m}
	+ \cdots + f_{AB_mB}\ \Omega _{B_1 \cdots B_{m-1} B}
\eqno (2.12)
$$
$L_A$ obeys the commutation relations of $\G$ of (2.2), vanishes on forms
of rank 0, and its square $L^2=L_AL_A$ is the quadratic Casimir operator
acting on the representation of $G$ under which the tensor form $\Omega
_{B_1\cdots B_m}$ transforms.

There are fundamental relation [18] between the actions of $D$, $i_A$ and
$L_A$ valid on tensor forms of any rank :
$$
\eqalign{
L_A i_B - i_B L_A = & f_{ABC}i_C \cr
 L_A D - D L_A = & 0 \cr
i_A D   + D i_A  = & - L_A
\cr}
\eqno (2.13)
$$
In particular, on tensor forms $\Omega $ of rank 0, the last identity reduces
to $i_A d \Omega + D i_A \Omega =0$.

Finally, we introduce a composite operation $\Delta$, which is neither a
derivative nor an anti-derivative, and is defined by
$$
\Delta \Omega _{B_1 \cdots B_m} = L_A (i_A \Omega _{B_1 \cdots B_m} )
\eqno (2.14)
$$
An analogous operation was introduced in [10] for the study of Lie algebra
cohomology. It obeys the following relations, valid on forms of any rank
$$
\eqalign{
L_A \Delta - \Delta L_A = & 0\cr
D\Delta  + \Delta D  = &  -L^2 \cr}
\eqno (2.15)
$$
The last equation will be a crucial ingredient in our analysis of cohomology.
It basically implies that any closed form which transforms under a
non-trivial representation of $\G$ is an exact form on $G$.

\bigskip

\noindent
{\it Outline of the Construction of Cohomology Generators of G/H}

\medskip

First, we use the result -- which was already mentioned [17,18] --
that the de Rham cohomology of $G/H$ can be identified with the coset
of the space of all closed $G$-invariant forms on $G/H$ by the space of
exterior derivatives of $G$-invariant forms on $G/H$.

Second, using Theorem 1, $G$-invariant forms on $G/H$ are given by
(2.4) where the coefficients $\omega _{A_1 \cdots A_n}$ are

{\it
\item{~~~~(1)} constant as functions on G,
\item{~~~~(2)} invariant under the adjoint action of $\H$, and
\item{~~~~(3)} zero whenever any of the indices $A_i$
corresponds to a generator of $\H$.
}

\noindent
Thus, all cohomology generators on $G/H$ are obtained as
forms of the type (2.4) that are closed and satisfy properties (1 - 3) modulo
differentials of forms that satisfy properties (1 - 3).
The space of forms obeying these properties is finite dimensional.

Third, we construct all closed forms obeying (1), including all exact forms
of the type (2.4), but ignoring properties (2) and (3) temporarily. To do
this, we make use in \S 3 of the operation $\Delta $ of (2.14) to map closed
forms onto closed forms of lower degree but higher rank. This map
is closely related to the transgression map [6,7] and produces a
{\it hierarchy} of equations which is closely related to the
descent equations. Ultimately, one is led to analyzing forms of degree 0 (i.e.
constants) which are $\G$-invariant completely symmetric tensors. This problem
can be studied using group theoretical methods alone. For example the analysis
of cohomology of degree 5 reduces to the study of constant symmetric tensors of
rank 3, also known as the $d$-symbols of $\G$.

Fourth, property (3) will in general put restrictions on the allowed
symmetrical tensors discussed in the third point above. In particular, the
$d$-symbols must vanish whenever all of its indices correspond to generators
of $\H$.

Fifth, one may integrate the hierarchy of equations and obtain all
closed forms of the type (2.4) that obey properties (1) and (2).
Property (3) is imposed explicitly on these forms and the resulting equations
can be solved rather easily. This gives all closed forms of the type (2.4)
obeying all three properties (1 - 3).

Finally, we discard exact generators
by simply enumerating the differentials of forms of the type (2.4) that obey
properties (1 - 3) themselves. The remaining forms are precisely
the de Rham cohomology generators of $G/H$.

\bigskip
\bigskip

\no
{\bf 3. Cohomology and Constant Invariant Tensors}

\medskip

In this section as well as in \S 4 and \S 5, we assume that $G$ is compact,
semi-simple, simply connected, and that $H$ is connected. The Lie
groups $G$ and $H$ admit the following factorization in terms of simple factors
$G_i$ and $H_j$ (with corresponding Lie algebras $\G _i$ and $\H _j$), and
$U(1)$ components :
$$
G =  G _1 \times \cdots \times G_p, \qquad\qquad
H =  H _1 \times \cdots \times H_q \times U(1)^r
\eqno(3.1)
$$
As a result, $G/H$ is simply connected and has vanishing first cohomology :
$H^1(G/H;R)=0$, which considerably simplifies the discussion.
These assumptions will be relaxed in \S 6, where the general case will be
considered.

We begin by formulating the correspondence that maps forms into forms of
lower degree but increased rank.  Let $\Omega$ be a closed $G$-invariant form
of rank 0 and degree $n$.   Using the third equation in (2.13) for forms of
rank 0, we see that
$$
d\Omega =0 \qquad
\Rightarrow \qquad D(i_A\Omega)=0
\eqno (3.2)
$$
We have thus constructed from $\Omega$ a new form $i_A \Omega$ which is
again closed. We show that $i_A\Omega$ is exact as a form on $G$  and we
have
$$
i_A \Omega =  D \Omega _A
\eqno (3.3)
$$
The form $\Omega _A$ may be constructed explicitly with the help of (2.15),
and the fact that the quadratic Casimir operator is invertible on
$i_A\Omega$, since the group $G$ is semi-simple.
Equation (3.3) determines $\Omega _A$ up to a closed form $\Omega '_A$.
Using (2.15) and the invertibility of the quadratic Casimir operator on
$\Omega '_A$, we find that $\Omega '_A$ is exact on $G$, so that $\Omega '_A
= DM _A$ for some form $M_A$ of degree 2 and rank 1. The addition of $\Omega
'_A$ does not modify the original forms $i_A \Omega $ or $\Omega$. Denoting by
$C_2(\G _j)\not=0$ the eigenvalue of the quadratic Casimir operator on each
simple component in the adjoint representation, we find
$$
\eqalign{
 \Omega _A = & DM _A  -{1 \over L^2} \Delta (i_A \Omega)
\cr
 = & DM _A - \sum _{j=1} ^p  {1\over C_2(\G_j)}\delta ^j _{AA'}
f_{A'BC}\ i_B i_C \Omega
\cr}
 \eqno (3.4)
$$
Here, $\delta ^j$ stands for the Kronecker symbol restricted to the
the simple component $\G _j$ of $\G$.
{}From $\Omega _A$, we construct a new closed form, again using (2.13)
$$
i_A \Omega = D \Omega _A \qquad  \Rightarrow \qquad
D(i_A \Omega _B + i_B \Omega _A ) =0
\eqno (3.5)
$$
The correspondence which maps the form $i_A \Omega$ onto the
form $i_A \Omega _B + i_B \Omega _A$ is the simplest case
of a {\it transgression map}, in which a closed tensor form of rank 1 and
degree $n-1$ is mapped onto a closed tensor form of rank 2 and of degree
$n-3$.

To generalize the above construction to tensor forms of arbitrary rank,
we have to overcome two complications. First, when an arbitrary  form $\Omega
_{B_1 \cdots B_m}$ is closed, the form $i_ B \Omega _{B_1 \cdots B_m }$ does
not, in general, obey any simple closure relation. Instead, the symmetrized
form $i_{\{ B} \Omega _{B_1 \cdots B_m \}}$, constructed from a tensor form
$\Omega _{B_1 \cdots B_m}$ which is itself symmetrical in all its $B_i$
indices, is automatically closed in view of (2.13).  Thus, for totally
symmetrical tensor forms   $\Omega _{B_1  \cdots B_m}$ we have the general
formula $$
D\Omega _{B_1 \cdots B_m} =0
\quad \Rightarrow \quad
D(i_{\{ B} \Omega _{B_1 \cdots B_m \} })=0
\eqno (3.6)
$$
Second, the tensor form $i_{\{ B} \Omega _{B_1 \cdots B_m \}}$ no longer
transforms under a single irreducible representation of $\G$, which
complicates the evaluation of the quadratic Casimir operator $L^2$ in
(2.15). Thus, we decompose the tensor into irreducible
components  $i_{\{B_{m+1}} \Omega _{B_1\cdots B_m\}}  (R_k)$ which transform
under the irreducible representation $R_k$ of $\G$
$$
i_{\{ B_{m+1}} \Omega _{B_1\cdots B_m\}} =\sum_{\{ R_k \} }
 i_{\{ B_{m+1}} \Omega _{B_1\cdots B_m\}}  (R_k)
\eqno (3.7)
$$
The quadratic Casimir operator within each irreducible representation $R_k$ is
just a numerical constant $C_2(R_k)$.

Using these ingredients, the above map may be generalized to act on
tensor forms of any rank and degree, which is the content of the following
Theorem. Let $\Omega _{B_1 \cdots B_m}$ be a closed tensor form of rank $m$ and
of degree $n\geq 2 $, which is completely symmetric in its indices $B_i$.

\medskip

\no
{\bf Theorem 2}

\medskip

{\it The forms $i_{\{ B_{m+1}} \Omega _{B_1\cdots B_m\}}$, and
every irreducible component $i_{\{ B_{m+1}} \Omega _{B_1\cdots B_m\}}
(R_k)$ defined in (3.7),  are automatically closed
$$
D\Omega _{B_1 \cdots B_m}=  0 \quad \Rightarrow \quad
D \bigl ( i_{\{ B_{m+1}} \Omega _{B_1\cdots B_m\}}  (R_k) \bigr ) =0
\eqno (3.8)
$$
For non-trivial irreducible representations $R_k$, the form
$i_{\{B_{m+1}}\Omega _{B_1 \cdots B_m \} } (R_k)$  is exact on $G$, and
determines a form $\Omega _{B_1 \cdots B_{m+1}}(R_k)$ up to an exact
form $DM _{B_1 \cdots B_{m+1}}(R_k)$ on $G$
$$
\eqalign{
 i_{\{ B_{m+1}} \Omega _{B_1\cdots B_m\}}  (R_k)
& = D\Omega _{B_1\cdots B_{m+1}} (R_k) \cr
&\cr
\Omega _{B_1\cdots B_{m+1}} (R_k)
 =   DM _{B_1 \cdots B_{m+1}} (R_k) &
-{m+1\over C_2(R_k) } f ^{AB} {}_{\{ B_{m+1}} i^A i_{\{ B}
\Omega _{B_1 \cdots B_m \} \} } (R_k) \cr}
\eqno (3.9)
$$
The form  $i_{\{B_{m+2}}\Omega _{B_1 \cdots B_{m+1} \} } (R_k)$ is
automatically closed in view of (2.13). For trivial representations $R_0$, the
form $i_{\{B_{m+1}}\Omega _{B_1 \cdots B_m \} } (R_0)$ is a
linear combination of products of $G$-invariant tensors and forms of rank 0
and degree $n-1$.}

\medskip

Theorem 2 is proved by repeated use of (2.13) and (2.15), and by exploiting
the fact that the quadratic Casimir operator is simply evaluated in any
irreducible representation of $G$.

The  map that sends a closed form  $i_{\{B_{m+1}}\Omega _{B_1 \cdots B_m \} }
(R_k)$  with non-trivial representation $R_k$ into another closed form
$i_{\{B_{m+2}}\Omega _{B_1\cdots B_{m+1} \} } (R_k)$, defined through Theorem
2 is essentially the transgression map of [6,7].  It reduces the degree of a
form by two units, while increasing the rank by  one unit, and may be iterated
until forms or degree 0 (or 1) are encountered. Since we assumed in this
section that $G/H$ is simply-connected, cohomology of degree 1 vanishes. The
study of invariant  forms of degree $n$ and rank 0 is thus reduced to the
analysis of constant invariant tensors.  This is indeed a  fundamental result
on cohomology of groups and homogeneous spaces [7]. The algebra of  tensors
$\omega _{B_1\cdots B_m;A_1\cdots A_n}$, completely  symmetrized in $B_i$ and
completely anti-symmetrized in $A_i$ is usually referred to as the Weil
algebra [6,7,11]. Theorem 2 implies that the cohomology of the Weil algebra is
located entirely in forms of degree 0, i.e. constants.  We shall now provide a
more detailed analysis of how the cohomologies of $G$ and $G/H$ are
constructed.

 The components in the trivial representation in
the decomposition of (3.7) are invariant under the adjoint action of $\G$, and
automatically closed.  They are linear combinations of products of
$G$-invariant tensors $d^{(k)} _{B_1 \cdots B_{m+1}}$ of degree 0 and closed
$G$-invariant forms $\Sigma _{n-2m-1}^{(k)}$ of degree $n-2m-1$ and rank 0.
Using the procedure of Theorem 2, we may write down explicitly a
{\it hierarchy} of equations that result from the successive application of the
above map :
$$
\eqalign{
	D(i_B \Omega ) =0
\quad \Rightarrow & \quad
i_B \Omega =D\Omega _B ~
+ \bigl ( \sum _k d _B ^{(k)} \Sigma ^{(k)} _{n-1}\bigr ) \cr
D(i_{\{B_2} \Omega _{B_1\}}) =0
\quad \Rightarrow & \quad
i_{\{B_2} \Omega _{B_1\}} = D\Omega _{B_1 B_2} ~+
\bigl ( \sum _k d_{B_1 B_2} ^{(k)} \Sigma ^{(k)} _{n-3} \bigr ) \cr
%D(i_{\{B_3} \Omega _{B_1B_2\}}) =0
%\quad \Rightarrow & \quad
%i_{\{B_3} \Omega _{B_1B_2\}} = D\Omega _{B_1B_2B_3} ~+
%\bigl ( \sum _k d_{B_1 B_2 B_3} ^{(k)} \Sigma ^{(k)} _{n-5} \bigr ) \cr
\cdots & \cr
D(i_{\{B_{m+1}} \Omega _{B_1 \cdots B_m\}}) =0
\quad \Rightarrow & \quad
i_{\{B_{m+1}} \Omega _{B_1 \cdots B_m \}}
\cr
& ~~~~~~~~~~~ = D\Omega _{B_1 \cdots B_{m+1}} ~
+\bigl (\sum _k d_{B_1 \cdots B_{m+1}} ^{(k)} \Sigma _{n-2m-1} ^{(k)}\bigr )
\cr}
\eqno (3.10)
$$
The hierarchy of (3.10) forms
the basis for the analysis of the structure of the cohomologies of $G$ and of
$G/H$ when combined with the expression of the forms
$\Omega _{B_1 \cdots B_{m+1}}$ given in (3.9). Its structure is related to
that of the descent equations [8,9], which proceed in opposite direction by
increasing the degrees in $\theta$ instead.

The sum terms on the right hand side of each of the above lines correspond to
the contribution
\footnote{$^\dagger$} { With the assumption that $G$ is semi-simple, no sum
term appears on the right hand side of the first line, since there can be no
invariant tensors transforming under the adjoint representation of $G$.}
in Theorem 2 of the trivial representations $R_0$.
They  have been put in parentheses for the following reason. If a term $\sum _k
d^{(k)} _{B_1 \cdots B_m} \Sigma ^{(k)} _{n-2m+1}$ arises in a given line $m$,
then a contribution from the hierarchy for the closed forms
$\Sigma ^{(k)} _{n-2m+1}$ must be subracted from the left hand side of the
equation of the next line $m+1$. This equation becomes instead
$$
D \bigl (i_{\{B_{m+1}} \Omega _{B_1 \cdots B_m\}}  -\sum _k d^{(k)} _{\{
B_1 \cdots B_m} i_{B_{m+1}\}} \Sigma ^{(k)} _{n-2m+1} \bigr ) =0
\eqno (3.11)
$$
The modifications required by (3.11) are understood in (3.10), but
have not been written explicitly to keep the notation as transparent
as possible.

The origin of the sum terms in (3.10) may be understood as follows.
Given cohomology generators $\Omega ^1$ and $\Omega ^2$ of degrees $n_1>0$ and
$n_2>0$, their (wedge) product $\Omega = \Omega ^1 \Omega ^2$ of degree
$n=n_1+n_2$ is a closed form and is a cohomology generator of degree
$n$. The hierarchy equations (3.10) for $\Omega$ may be simply deduced from
those for $\Omega ^1$ and $\Omega ^2$ and will contain sum terms on the right
hand side of (3.10) with $\Sigma$ of degree $n_1$ or $n_2$. More generally,
sum terms on the right hand side of (3.10) occur with a degree of $\Sigma$
greater or equal to 1 if and only if the form $\Omega$ contains linear
combinations of products of cohomology generators of strictly lower degree.

It is standard to introduce the notion of {\it primitive generators}, which do
not contain any linear combinations of products of generators of strictly
lower degrees. Generators that do contain linear combinations of products of
cohomology generators of strictly lower degree are said to be {\it
decomposable generators}. Clearly, the cohomology is completely determined by
its primitive generators, since decomposable generators may always be obtained
as linear combinations of products of primitive generators including those of
lower degree. Henceforth, without loss of generality, we limit ourselves to the
analysis of the primitive cohomology generators.

If $\Omega$ is a {\it primitive generator} of degree $n$, the associated
hierarchy contains no $\Sigma ^{(k)}$ sum terms on the right hand side of
(3.10), except possibly when their degree vanishes.
The hierarchy of (3.10) then considerably
simplifies, and terminates at $m= [(n-1)/2]$ where forms of degree 0 are
encountered. No subtractions of the type encountered in (3.11) are required
in this case and we have
$$
\eqalign{
	D(i_B \Omega ) =0
\quad \Rightarrow & \quad
i_B \Omega =D\Omega _B \cr
D(i_{\{B_2} \Omega _{B_1\}} )=0
\quad \Rightarrow & \quad
i_{\{B_2} \Omega _{B_1\}} = D\Omega _{B_1 B_2} \cr
D(i_{\{B_3} \Omega _{B_1B_2\}} )=0
\quad \Rightarrow & \quad
i_{\{B_3} \Omega _{B_1B_2\}} = D\Omega _{B_1B_2B_3} \cr
\cdots & \cr
D(i_{\{B_{m+1}} \Omega _{B_1 \cdots B_m\}} )=0
\quad \Rightarrow & \quad
i_{\{B_{m+1}} \Omega _{B_1 \cdots B_m \}} = D\Omega _{B_1 \cdots B_{m+1}} +
 d _{B_1 \cdots B_{m+1}}
\cr}
\eqno (3.12)
$$
Here, $d _{B_1 \cdots B_{m+1}}$ is a constant $G$-invariant tensor, completely
symmetric in its indices; when $i_{\{ B_{m+1}} \Omega _{B_1 \cdots B_m}$
above is of degree 0, we set $\Omega _{B_1 \cdots B_{m+1}}=0$.
The hierarchy (3.12) associates to a primitive
generator $\Omega$ a sequence of forms that terminates with a constant
symmetric $G$-invariant tensor.
The analysis of cohomology is thus reduced to the analysis of
constant tensors with certain invariance properties, a problem that can be
solved by group theoretical methods alone.

The cases for $n$ odd or even differ substantially, and are best analyzed
separately. When $n=2m+1$ is odd, the last equation in (3.12) simplifies, as
$\Omega _{B_1 \cdots B_{m+1}}=0$ and we find that to each primitive
generator of odd degree there corresponds a constant $G$-invariant completely
symmetric tensor $d _{B_1 \cdots B_{m+1}}$ of rank $m+1= (n+1)/2$.
Conversely, a tensor $d _{B_1 \cdots B_{m+1}}$ will correspond to a primitive
generator on $G/H$ only provided it vanishes on $\H$. When $H=1$, we
have the cohomology of Lie groups, and no extra conditions are
needed. In this case, the invariant tensor $d _{B_1  \cdots B_{m+1}}$ produces
a unique cohomology generator of degree $2m+1$ on $G$, given by a well-known
expression  [8] :
$$
\Omega ^{(d)} \sim d _{B_1 \cdots B_{m+1}} \theta ^{B_1}
d\theta ^{B_2} \cdots  d\theta ^{B_{m+1}}
\eqno (3.13)
$$
$\G$-invariance of $d$ guarantees closure of $\Omega ^{(d)}$.

When $n=2m+2$ is even, the tensor $\Omega _{B_1 \cdots B_{m+1}}$ is of degree
0 and thus constant. In contrast with the case of $n$ odd, it does not have to
be $\G$-invariant, but should be invariant only under the action of $\H$ in
the adjoint representation. Thus, even dimensional cohomology of $G/H$ is
related to constant $\H$-invariant tensors, again with certain vanishing
conditions on $\H$.

\bigskip
\bigskip

\no
{\bf 4. Low Dimensional Cohomology of degrees 2, 3 and 4}

\medskip

In this section, we compute explicitly all generators of low
dimensional cohomology using the methods of \S 3. We make the same
assumptions : $G$ and $H$ are compact  and connected, $G$ is
semi-simple and simply connected and  $H$ is semi-simple times $U(1)$ factors
as
in (3.1), so that $H^1(G/H;R)=0$. We shall also re-express our results
directly in terms of the Goldstone field $U$, the composite $\H$-valued gauge
field $V$ with components $\theta ^\alpha$, its field strength $W$ and
associated covariant derivative $D_\H$, which are defined as follows
$$
V=\bigl ( U^{-1} dU \bigr ) _\H
\qquad \quad
W= dV + V^2
\qquad \quad
U^{-1} D_\H U = \bigl ( U^{-1} dU \bigr ) _\M
\eqno (4.1)
$$
In particular, this notation will allow us to make contact with the results
of [5].

\medskip

\no
{\it Cohomology of degree 2}

\medskip

We consider a closed form $\Omega$ of rank 0 and degree 2.
$$
\Omega = \half \omega _{a_1a_2} \theta ^{a_1} \theta ^{a_2}
\eqno (4.2)
$$
where $\omega _{a_1a_2}$ is $\H$-invariant and the notation of (4.2) shows
that the form $\Omega$ vanishes on $\H$ as required by Theorem 1. We follow
the procedure of Theorem 2, and associate to $\Omega $ a constant form $\Omega
_A$ of degree 0 and of rank 1 defined by (3.3--4).  The contribution from the
exact form $DM _A$ in (3.4) is absent and $\H$-invariance of $\omega _{ab}$
implies $\H$-invariance of $\Omega _A$. There are two possible cases for the
constant $\Omega _A$ :

1. $A=\alpha$, corresponding to a generator of $\H$.  If $A=\alpha$
corresponds to a generator in one of the simple components of $H$ in (3.1),
then the tensor $\Omega _\alpha$ cannot be $\H$-invariant.
If $\alpha$ corresponds to a generator of
one of the $U(1)$ factors in (3.1), then $\Omega _\alpha$ is invariant under
$\H$ and may just be set to a constant. Thus, for every $U(1)$ factor of $H$,
there is a generator of $H^2(G/H;R)$  given by
$$
\Omega ^{(l)} = \half \Omega _{\alpha _l} f_{\alpha _lab} \theta ^a \theta ^b
\qquad l=1,\cdots,r
\eqno (4.3)
$$
The forms $\Omega ^{(l)}$ are curvature forms of composite Abelian
connections $V$ of (4.1) associated with each of the commuting $U(1)$ factors
in $H$
$$
\Omega ^{(l)} = -d (\Omega _{\alpha _l} \theta ^{\alpha _l})
=-\Omega _{\alpha _l} W_{\alpha _l}
\eqno (4.4)
$$
These forms are the generators of the first Chern class and the associated
invariant effective action [5] is that of a charged particle moving in the
presence of a magnetic monopole~[4].

 2. $A=a$, corresponding to a generator of $\M$. This can occur if $\M$
contains a generator that commutes with all of $\H$, so that $f_{\beta aC}=0$
for all $\beta \in \H$. From $\Omega _a$, we obtain $i_B \Omega = D\Omega _B$,
and from $i_B \Omega$, we have
$$
\Omega
=\half f_{abc} \Omega _a \theta ^b \theta ^c
= - d(\Omega _a \theta ^a)
\eqno (4.5)
$$
The forms $\Omega$ are well-defined on $G/H$, vanish on $\H$, and are the
exterior derivatives of well-defined $\H$-invariant forms on $G/H$. These forms
are exact on $G/H$,  and do not contribute to $H^2(G/H;R)$.

\medskip

\no
{\it Cohomology of degree 3}

\medskip

We consider a closed form  $\Omega$ of rank 0 and of degree 3,
$$
\Omega = {1 \over 3!} \omega _{a_1 a_2 a_3} \theta ^{a_1} \theta ^{a_2} \theta
^{a_3}
\eqno (4.6)
$$
where $\omega _{a_1a_2a_3}$ is $\H$-invariant and vanishes on $\H$. Following
the procedure of Theorem 2, and using (3.3-4) we associate to $\Omega$ a form
$\Omega _A$ of degree 1 and an arbitrary form $M _A$ of degree 0.
Furthermore,  $d_{BC} = i_{\{ B} \Omega _{C \}} $ is closed and of
degree 0, which implies that it is a constant symmetric $G$-invariant tensor.
Within each simple component of $G$, $d_{AB}$ must be proportional to the
$G$-invariant metric $\delta ^j_{AB}$  on $\G_j$, so that we have
$$
d _{BC} = \sum _{j=1} ^p d _2 ^j\ \delta ^j _{BC}
\eqno (4.7)
$$
where $d_2^j$ are constants. From (3.4), it is clear that up to an exact form,
$\Omega _A$ must be independent of $\theta ^\beta$, and hence $d_{BC}$ must
vanish on $\H$ : $d_{\beta \gamma}=0$.

Conversely, to find all primitive closed forms $\Omega$ on $G/H$, we start from
the most general invariant tensor $d_{AB}$ of (4.7) satisfying $d_{\beta
\gamma}=0$ for all $\beta,~\gamma$ corresponding to generators in $\H$,
integrate the hierarchy equations (3.12) and enforce $\H$-invariance and
vanishing on $\H$, as required by Theorem 1. The most general solution is
obtained from $i_B \Omega _C$ by adding an arbitrary  constant anti-symmetric
tensor :
$$
i_B \Omega _C =
d _{BC} + K_{BC} \qquad K_{BC}= - K_{CB}
\eqno (4.8)
$$
The contribution of $M _A$ in (3.4) can always be absorbed into $K_{AB}$.
Independence of $\Omega _C$ on $\theta ^\beta$ requires $d_{\beta c} =
-K_{\beta c}$ and $K_{\beta \gamma}=0$. The contribution to
$\Omega $ of the $K_{ab}$ components of $K$ is always exact, and will be
omitted. The form $\Omega $ is easily reconstructed and  we find
$$
\Omega =
{1 \over 6 } d _{ab}f_{bcd}\ \theta ^a \theta ^b \theta ^c +
{2 \over 3 } d _{a\beta}f_{\beta cd}\ \theta ^a \theta ^b \theta ^c
\eqno (4.9)
$$
Notice that the components $d_{ab}$ and $d_{a \beta}$ are related by
$\G$-invariance of $d_{AB}$.

 To construct the form $\Omega$ from $d_{AB}$, one
may also proceed directly from the familiar Goldstone-Wilczek current, (which
is in the same as the WZW term in 1+1 dimensions) in which the subgroup $H$ has
been gauged [12,15,19].  The precise correspondence is achieved by identifying
the invariant metrics on $\G_j$ with the Cartan-Killing form $\delta ^j _{BC} =
\tr ~T^j_B T^j _C=\tr _{\G _j} T_B T_C$ and making use of the notation of
(4.1). Here $\tr _{\G _j}$ stands for the trace in the simple component
$\G_j$ only. It is easy to see that $\Omega$ of (4.9) is then given by a
sum of gauged Goldstone-Wilczek forms
$$
\Omega = { 1 \over 3} \sum _j d_2 ^j ~\tr _{\G_j} \biggl \{
\bigl ( U^{-1}D_\H U \bigr ) ^3 - 3 W U^{-1} D_\H U \biggr \}
\eqno (4.10)
$$
The exterior derivative of $\Omega$ vanishes in view of the fact that
$d_{\alpha \beta}=0$ for $\alpha$ and $\beta$ corresponding to any generators
of $\H$. This form is of course nothing but the gauged WZW term in 2
dimensions for the coset model $G/H$.

 Since we have
assumed that $H^1(G/H;R)=0$, there are no decomposable generators of degree
3, and third cohomology is completely characterized by symmetric,
$\G$-invariant, rank 2 tensors on $\G$ that vanish on $\H$.

\medskip

\no
{\it Cohomology of degree 4}

\medskip

Finally, we consider a form $\Omega$ of degree 4,
$$
\Omega = {1 \over 4!} \omega _{a_1 \cdots a_4} ~\theta ^{a_1} \cdots
\theta ^{a_4}
\eqno (4.11)
$$
where $\omega _{A_1 \cdots A_4}$ is $\H$-invariant and vanishes on $\H$.
{}From the closure of $\Omega$ and (3.3-4), we associate to $\Omega$ the
form $\Omega _A$ of degree 2, and an arbitrary form $M _A$ of degree 1.
Using (3.5), we see that the form $i_{\{ A} \Omega _{B\} }$ is closed, and we
again apply Theorem 2 to it. Since $i_{\{ A} \Omega _{B\}}$ is a one form it
must be a linear combination of the left-invariant forms $\theta ^A$. Thus,
upon
decomposing $i_{\{ A} \Omega _{B\}}$ into irreducible representations of $\G$,
we only encounter the adjoint representation. (If $\G$ had not been
semi-simple, we would also have encountered trivial representations at this
point, associated with possible $U(1)$ components of $G$.) As a result, $i_{\{
A} \Omega _{B\}}$ is an exact form according to Theorem 2, and we have
$$
i_{\{ A} \Omega _{B\} } = D\Omega _{AB}
\eqno (4.12)
$$
The form $\Omega _{AB}$ is of degree 0 and of rank 2, symmetric in $A$ and
$B$ and must be invariant under the adjoint action of $\H$.

To find all primitive closed forms $\Omega$, we proceed in analogy with the
case of degree 3 : we integrate the hierarchy equations (3.12) and impose the
conditions that the forms should vanish on $\H$ and be $\H$-invariant.
The most general solution to (4.12) is obtained by adding to $D\Omega _{BC}$
an arbitrary  anti-symmetric form $K$ of degree 1 and rank 2
$$
i_B \Omega _C = D \Omega _{BC} + K_{BC} \qquad \qquad K_{BC}=-K_{CB}=
k_{BC;A} \theta ^A
\eqno (4.13)
$$
Here, $K_{BC}$ and $\Omega _{BC}$ are  constrained only
by the requirement that (4.12) be integrable as an equation in $\theta ^A$.
The most general solution is easily found and we have
$$
\Omega _A =  -D(\Omega _{AB} \theta ^B) - \half \tilde k _{ABC} \theta ^B
\theta ^C
\eqno (4.14)
$$
Here, $\tilde k _{ABC}$ is an arbitrary $\H$-invariant tensor of rank 3 which
is completely anti-symmetric in its three indices. Because the
contribution to $\Omega _A$ involving $\Omega _{BC}$ in (4.14) is
an exact differential, it drops out in the evaluation of $i_A \Omega$, and
we find that
$$
\Omega =  d Q \qquad \qquad Q = { 1 \over 3!} \tilde k _{ABC} \theta ^A
           \theta ^B \theta ^C
\eqno (4.15)
$$
To render this form well-defined on $G/H$, we require that
$\Omega$ be independent of any of the $\theta ^\lambda$, where $\lambda$
corresponds to a generator in $\H$. Putting this requirement
directly on $\Omega$ turns out to be rather cumbersome; it is more
convenient to require that the combination $\Omega _A - DM _A$ of (3.4) and
(3.9) be independent of $\theta ^\lambda$.
$$
i_\lambda (\half \tilde k _{ABC} \theta ^B \theta ^C + D\tilde M _A) =0
\eqno (4.16)
$$
This equation is easy to solve in terms of a constant tensor $m_{AB}$ with
$$
\tilde k _{A\lambda B} = m_{C \lambda} f_{ABC} \qquad \qquad \tilde M _A =
M _A + \Omega _{AB} \theta ^B =
m_{AB} \theta ^B
\eqno (4.17)
$$
Addition of an arbitrary exact form to $Q$ leaves $\Omega $ unchanged, and is
equivalent to shifting $m_{AB}$ by an arbitrary anti-symmetric rank 2 tensor.
Using this property, we may chose $m_{\alpha \beta}$ symmetric and $m_{a
\beta}=0$. Neglecting the contribution to $Q$ of well-defined forms on $G/H$,
which lead to exact contributions to $\Omega $ only, we find
 $$
Q=  \half m_{\beta \lambda } f_{a \beta c} \theta ^a \theta ^c \theta ^\lambda
    +{ 1 \over 6} m_{\beta \lambda } f_{\alpha \beta \gamma} \theta ^\alpha
\theta ^\gamma \theta ^\lambda
\eqno (4.18)
$$
which gives the following expression for $\Omega$
$$
\Omega = m_{\alpha \beta} W_\alpha W_\beta \qquad \qquad
W_\alpha = d\theta _\alpha + \half f_{\alpha \gamma \delta} \theta _\gamma
\theta _\delta
\eqno (4.19)
$$
Here, $W_\alpha$ are the components of the curvature form $W$ of (4.1); and
$m_{\alpha \beta}$ is any constant symmetric tensor invariant under the adjoint
action of $\H$.
All such tensors may be parametrized in terms of the Cartan-Killing forms on
the simple components of $\H$, and arbitrary coefficients on the $U(1)$
components of $\H$ of (3.1). As a result, we have
$$
\Omega = \sum _{k=1} ^q m_2 ^{(k)} W_\alpha ^{(k)} W_\alpha ^{(k)}
+ \sum _{l,m=1} ^r m_1 ^{(l,m)} W_{\alpha _l} W_{\alpha _m}
\eqno (4.20)
$$
The generators of $H^4 (G/H;R)$ in the first sum belong to the second Chern
class (or first Pontryagin class) evaluated on the $\H$-valued connection
$V$ of (4.10) with components $\theta ^\alpha$, while the generators in
the second sum arise from products of generators belonging to the  first Chern
class. The latter were already identified completely in (4.4) in the
subsection on cohomology of degree 2. Since $H^1(G/H;R)=0$, there are no
further decomposable generators. The form $Q$ in (4.18) may be viewed as a
linear combination of Chern-Simons actions in 3 dimensions [8,13], evaluated on
the composite gauge field $V$ of (4.1). The resulting invariant effective
action [5] thus coincides with the Chern-Simons action evaluated on composite
connections and provides generalizations of the Hopf invariant. (see for
example [20] for general references)

\bigskip
\bigskip

\no
{\bf 5. Cohomology of Degree 5 and the G-Invariant $d$-Symbol}

\medskip

Let $\Omega$ be a closed form of rank 0 and degree 5 on $G/H$ given
by
$$
\Omega = {1 \over 5!}\ \omega _{a_1 \cdots a_5} \theta ^{a_1} \cdots
\theta ^{a_5}
\eqno (5.1)
$$
Following the procedure of Theorem 2, and equations (3.12)
for primitive cohomology generators, we associate to $\Omega $ the hierarchy
$$
\eqalign{
&i_A \Omega =  D\Omega _A
{}~~~~~~~\qquad \Rightarrow \qquad
\Omega _A = DM _A - ( L^2 )^{-1} \delta ( i_A \Omega )
\cr
&i_{\{ A } \Omega _{B\}} =  D\Omega _{AB}
\qquad \Rightarrow \qquad
\Omega _{AB} = DM _{AB} - ( L^2)^{-1} \delta ( i_{\{A } \Omega _{B\}}
)  \cr
&i_{\{ A} \Omega _{BC\}}=  d_{ABC}
\cr}
\eqno (5.2)
$$
Here, $d_{ABC}$ is a constant totally symmetric and $G$-invariant tensor on
$\G$, referred to as the $d$-symbol of $G$ in the physics literature on chiral
anomalies [14]. It is a simple group theoretical problem to find all such
invariant tensors. Assuming that $\G$ is semi-simple, the most general
$d$-symbol on $\G$ is parametrized completely by the $d$-symbols on the
simple components $\G_j$.
$$
d_{ABC} = \sum _{j=1} ^p d _3 ^j \ d^j _{ABC}
\eqno (5.3)
$$
On each simple component $\G_j$, there is at most one such invariant tensor.
For all simple compact Lie groups, we have $d_{ABC}=0$, except for $SU(N)$,
$N\geq 3$, where $d_{ABC}\not=0$; this includes the case of $SO(6)=SU(4)/Z_2$
[14].

In the simplest case where $H=1$, and $G/H$ is a Lie group, there are no
further conditions on the $d$-symbols and we shall see that each non-zero
$d$-symbol produces a non-trivial cohomology generator of $G$, given
by (3.13).

For more general subgroups $H$, the $d$-symbols corresponding to
cohomology generators on $G/H$ must satisfy further conditions.
A complete characterization of $H^5(G/H;R)$ is given by the following
Theorem~:

\medskip

\no
{\bf Theorem 3}

\medskip

{\it
The $d_{ABC}$ tensor associated with a closed 5-form $\Omega$ on $G/H$ must
vanish on the subalgebra $\H$ (i.e. $H$ must be embedded into $G$ in
an anomaly free way). Conversely, any non-zero $G$-invariant tensor $d_{ABC}$,
which vanishes on the subalgebra $\H$ produces a unique primitive generator of
$H^5(G/H;R)$. All other generators of $H^5(G/H;R)$ are decomposable into
linear combinations of products of generators of degrees 2 and 3.
 }

\medskip

To show that a $d$-symbol corresponding to a cohomology generator on
$G/H$ must vanish on $\H$, we produce an explicit formula for $d_{ABC}$ in
terms of $\Omega$ of (5.1) by solving (5.2). The exact forms $DM _A$ and
$DM _{AB}$ of (5.2) do not contribute to $d_{ABC}$ : $DM _{AB}$ would add
to $d_{ABC}$ a quantity $-Di_{\{ A} M _{BC\}}$ which vanishes since $M
_{BC}$ is of degree 0; $DM _A$ contributes $-i_{\{ A} M _{B\}}$
to $\Omega _{AB}$ which would add to $d_{ABC}$ a quantity $-i_{\{ A}
i_{\{ B} M _{C \} \} }=0$. It follows that the $d$-symbol is given in terms
of $\Omega$ by the following expression.
$$
d_{ABC} = i_{\{ A} {1 \over L^2} \delta \biggl (
i_{\{ B} {1 \over L^2} \delta \bigl ( i_{C\}\}} \Omega \bigr ) \biggr )
\eqno (5.4)
$$
Both Casimir operators effectively act on the adjoint representation of $\G$,
and (5.4) may be evaluated in terms of the symbols $f^j _{ABC}$ which are the
structure constants of $\G$, restricted to the simple component $\G_j$,
and we find
$$
d_{ABC} = \sum _j {1 \over 3! ~ C_2 (\G _j ) ^2 }
f^j _{BA_2 A_3} f^j _{C A_4 A_5} \omega _{A A_2 A_3 A_4 A_5}
\eqno (5.5)
$$
The expression on the right hand side is completely symmetrized in $ABC$. Also,
one of the three indices $A,~B,~C$  is always attached directly to $\omega$.
Since $\omega$ vanishes whenever one of its indices corresponds to a generator
in $\H$, the $d$-symbol restricted to the subalgebra $\H$ must vanish :
$d_{\alpha \beta \gamma}=0$, for all indices $\alpha,~\beta, ~\gamma$
corresponding to generators of $\H$. This is precisely the content of the
first part of Theorem 3.

To show that a non-zero $G$-invariant tensor $d_{ABC}$, which vanishes on
the subalgebra $\H$, produces a unique primitive generator of $H^5 (G/H;R)$,
we integrate the hierarchy equations of (5.2).
The first step is to obtain the one form $\Omega _{BC}$ from (5.2);
it is determined up to tensors $ l_{BC;A} = l_{CB;A} $ as follows
$$
\Omega _{BC} = ( d_{ABC} + l _{BC;A} ) \theta ^A
\qquad \qquad
l _{BC;A} + l _{AB;C} + l _{CA;B}=0
\eqno (5.6)
$$
Applying the differential $D$ to the above expression for
$\Omega _{BC}$, we find $i_B \Omega _C + i_C \Omega _B$, which
determines $\Omega _A$ up to an anti-symmetric form $K$ :
$$
i_A \Omega _B = D\Omega _{AB} + K_{AB} \qquad K_{AB} = - K_{BA} =\half
k_{AB;CD} \theta ^C \theta ^D
\eqno (5.7)
$$
{}From $\Omega_A$, we have $i_A\Omega = D\Omega _A$, and from  $i_A
\Omega$, we obtain the general solution for $\Omega$.
$$
\Omega = \Omega ^{(d)} + \Omega ^{(k)} + \Omega ^{(l)}
\eqno (5.8)
$$
The three contributions arise from the $d$ tensor and from the $k$
and $l$ forms respectively, and are given by
$$
\eqalign{
\Omega ^{(d)} &= {1 \over 40}\ d_{A_1BC}\ f_{BA_2A_3}\ f_{CA_4A_5}\
\theta ^{A_1} \cdots \theta ^{A_5}
\cr
\Omega ^{(k)}       &= d Q ^{(k)}
\qquad \qquad
Q ^{(k)} = {1 \over 24} k_{A_1A_2;A_3A_4}\ \theta ^{A_1} \cdots \theta ^{A_4}
\cr
\Omega ^{(l)} &= d Q^{(l)}
\qquad \qquad
Q^{(l)} = {1 \over 12} l _{A_1M;A_2}\    f_{MA_3A_4}\ \theta ^{A_1} \cdots
\theta ^{A_4}
\cr}
\eqno (5.9)
$$

In the case where $H=1$, there are no further conditions on the forms $\Omega
^{(d)}$, $\Omega ^{(k)}$ and $\Omega ^{(l)}$, which are well-defined on $G$.
The forms $\Omega ^{(k)}$ and $\Omega ^{(l)}$ are exact on $G$, while for
$d_{ABC}\not=0$, the form $\Omega ^{(d)}$ produces a non-trivial generator of
$H^5(G;R)$. For simple compact groups, this cohomology has a single generator
for $SU(N)$, $N \geq 3$ and $SO(6)=SU(4)/Z_2$, while it is trivial for all
other cases [21]. For products of groups, one makes use of the K\"unneth
formula, as in [5].

A more familiar expression for the form $\Omega ^{(d)}$ is obtained by
casting $d_{ABC}$ in terms of a trace over representation matrices $T_A^j$ of
the simple components $\G _j$ of $\G$ : $ d ^j _{ABC}= \tr~ T^j _A T^j
_{\{ B} T^j _{C\}} = \tr _{\G _j} T_A T_{\{ B} T_{C\}} $.
Substituting into $\Omega ^{(d)}$, and using the fact that the
structure constants in (5.9) yield commutators, we obtain
$$
\Omega ^{(d)} = \sum _j  {1 \over 10} d^j _3 ~\tr _{\G _j} \bigl ( U ^{-1}
dU \bigr ) ^5
\eqno (5.10)
$$
The forms $\tr _{\G _j} (U^{-1} dU )^5$
are the generators of $H^5(G;R)$ for each
simple component and correspond to the standard WZW terms on $G_j$.

For more general groups $H$, it is still the case that each non-zero
$d$-symbol (which vanishes on $\H$, according to the first part of Theorem 3)
gives rise to a non-trivial cohomology generator in $H^5(G/H;R)$. To see
this, we must add to $\Omega ^{(d)}$ contributions of the form $\Omega ^{(k)}$
and $\Omega ^{(l)}$ to guarantee that the resulting form will properly vanish
on $\H$, and can thus be projected down to $G/H$.
The final result is a form $\tilde \Omega ^{(d)}$ which vanishes on $\H$ and
is given by
$$
\eqalign{
\tilde \Omega ^{(d)} = & {1 \over 40} \bigl \{
d_{a_1 bc} f_{ba_2a_3} f_{ca_4a_5}
+ 7 d_{a_1 b\gamma } f_{ba_2a_3} f_{\gamma a_4a_5} \cr
& + 16 d_{a_1 \beta \gamma } f_{\beta a_2a_3} f_{\gamma a_4a_5} \bigr \}
\theta ^{a_1}\theta ^{a_2}\theta ^{a_3}\theta ^{a_4}\theta ^{a_5}
\cr}
\eqno (5.11)
$$
Notice that $d_{\alpha \beta \gamma}$ does not enter, and that for $H=1$, we
recover $\Omega ^{(d)}$ of (5.9).

An alternative procedure for obtaining the same result is
already quite familiar and was used in [5]. In terms of the
$\H$-valued gauge field $V$, the $\H$-covariant derivative $D_{\H}$ of
(4.1), and the trace expression for the form $\Omega ^{(d)}$ of (5.10), we
obtain a form [9,22] that vanishes on $\H$
$$
\tilde \Omega ^{(d)} = \sum _j {1 \over 10}~ d_3 ^j ~\tr _{\G_j} ~
\biggl \{ \bigl ( U ^{-1} D_{\H} U \bigr ) ^5 - 5 W \bigl ( U ^{-1}
D_{\H} U \bigr )^3 +10 W ^2 \bigl (U ^{-1} D_{\H} U \bigr )\biggr \}
\eqno (5.12)
$$
As is well-known [5,14], closure of this form is guaranteed by the fact that
$d_{ABC}$ vanishes on $\H$. The generator obtained in this way cannot be
decomposed into a sum of products of generators of lower degree that are
well-defined on $G/H$, and thus  $\tilde \Omega ^{(d)}$ is {\it primitive}.

The last issue we must address is whether the remaining forms $\Omega ^{(k)}$
and $\Omega ^{(l)}$ -- which are exact on $G$ -- by themselves produce any
non-trivial cohomology generators on $G/H$. Actually, inspection of (5.9)
reveals that the forms $\Omega ^{(l)}$ are a special case of the forms
$\Omega ^{(k)}$, so that we restrict to the latter. We now impose
$\H$-invariance, vanishing on $\H$ and discard generators that are exact on
$G/H$.

{}From (3.4), we know that, up to an exact form on $G$, the form $\Omega ^{(k)}
_B$ must vanish on $\H$. Furthermore, because $\Omega ^{(k)}$ in (5.9) is
exact on $G$, the form $\Omega ^{(k)} _B$ equals $i_B Q^{(k)}$, up to an exact
form on $G$. Combining both, the condition for the vanishing of
$\Omega^{(k)}$ on $\H$ is that
$$
i_\alpha (i_B Q^{(k)} - DM_B) =0
\eqno (5.13)
$$
Using (5.13) and $\H$-invariance of the tensor $k_{A_1 \cdots
A_4}$, certain  components of this tensor can be expressed in terms of the form
$M_A=\half m_{A;BC} \theta ^B \theta ^C$
$$
k_{\alpha B;CD} =
m_{B;E\alpha}\ f_{ECD}
																		+m_{E;C\alpha}\ f_{EBD}
                  +m_{E;D\alpha}\ f_{ECB}
\eqno (5.14)
$$
Using (5.14), anti-symmetry of $k$ and $\H$-invariance of $m_{A;BC}$,  it is
easy to establish that $m$ must satisfy the following condition
$$
(m_{A;\epsilon C} + m_{C;\epsilon A })\ f_{\beta  \delta \epsilon } =0
\eqno (5.15)
$$
for all indices $A,C$ and $\delta, \ \beta$.
 Finally, the form $Q^{(k)}$ is defined only up to an
exact form on $G$. The addition of an arbitrary exact form $N= {1 \over 6}
n_{ABC} \theta ^A \theta ^B \theta ^C$ to $Q^{(k)}$ is equivalent to shifting
the form $M_B$ by $i_B N$, and amounts to shifting the tensor $m_{A;BC}$ by a
totally anti-symmetric tensor $n_{ABC}$. This freedom to shift $m$ may be
used to render it symmetric in its indices $A$ and $B$ for example.

Condition
(5.15) implies that for all indices corresponding to generators of $\H$, we
must have $k_{\alpha \beta;\gamma \delta}=0$. Condition (5.14) and
invariance of $k$ under the adjoint action of $\H$ furthermore imply that
the form $\Omega ^{(k)}$ of (5.9) is given by
$$
\Omega ^{(k)} = d Q^{(k)}
\qquad \quad
Q ^{(k)} = {1 \over 6} k_{\alpha b;cd}\ \theta ^\alpha \theta ^b \theta ^c
\theta ^d
\eqno (5.16)
$$
Here, the indices $\alpha$ only run over generators of $\H$ that commute with
all of $\H$.
These generators are in general non-trivial, because $\theta ^\alpha$ is not a
well-defined differential on $G/H$. Only its exterior derivative $d\theta
^\alpha = - \half f_{\alpha bc} \theta ^b \theta ^c$ is well-defined, and is a
closed 2 form on $G/H$ as shown in (4.4). The remaining form of degree 3
given by ${1 \over 6} k_{\alpha b;cd} \theta ^b \theta ^c \theta ^d$ is closed
on $G/H$ in view of (5.14) and corresponds to a non-trivial cohomology
generator of $H^3 (G/H;R)$.  Thus, the contributions to $\Omega^{(k)}$ are
linear combinations of forms that factorize into products of cohomology
generators of degrees 2 and 3. This concludes the proof of Theorem 3, and
the complete description  of cohomology of degree 5.

\bigskip
\bigskip

\no
{\bf 6. Cohomology of non-simply Connected G/H}

\medskip

We now generalize our results to the case where $G/H$ is not necessarily
simply-connected. This may occur because $G$ is not simply connected or
because $H$ is not connected.
$H^1(G/H;R)$ may then contain non-trivial generators, which
are closed forms of degree 1, invariant under the adoint action of $\H$.
Let us denote a set of independent generators by $\theta ^{s_i}$ with $i=1,
\cdots , t={\rm dim}~ H^1(G/H;R)$. The analysis of the
cohomology of degree $n$ proceeds directly from Theorem 1. Let $\Omega $ be a
closed $\H$-invariant form on $G/H$.
 $$
\Omega = {1 \over n!} \omega _{a_1 \cdots a_n} \theta ^{a_1} \cdots
\theta ^{a_n}
\eqno (6.1)
$$
Since the generators $\theta ^{s_i}$ are $\H$-invariant by themselves,
$\Omega$ may be written as a linear combination of forms of degree $n$ that
involve $k$ factors of generators $\theta ^{s_i}$ with $k\leq t$
$$
\Omega = \Omega ^{(n)} + \sum _{k=1} ^t \theta ^{s_{i_1}} \cdots
\theta ^{s_{i_k}}  \Omega ^{(n-k)} _{s_{i_1} \cdots s_{i_k}}
\eqno (6.2)
$$
Each of the forms $\Omega ^{(n-k)} _{s_{i_1} \cdots s_{i_k}}$ is closed,
invariant under the adjoint action of $\H$ and well-defined on $G/H$.  As a
result, each $\Omega ^{(n-k)}$ is just a generator of the cohomology group
$H^{n-k}(G/H;R)$ for $k \geq 1$, which does not involve any generators of the
first cohomology  group. Thus, the analysis of the forms $\Omega ^{(n-k)}$ is
just that of the cohomology for which no generators of degree 1 occur and has
already been carried out previously.

To summarize, when $H^1 (G/H;R ) \not=0$, the cohomology involving those
generators is formed out of the product of cohomology generators of
lower degree with products of cohomology generators of degree 1.

\bigskip
\bigskip

\noindent
{\bf 7. Gauging Invariant Effective Actions}

\medskip

So far we have dealt with effective actions for Goldstone boson fields only.
To couple the Goldstone fields to gauge fields, we follow the procedure of
minimal coupling, and gauge the global symmetry group $G$. Manifestly
invariant actions, corresponding to $G$-invariant Lagrangian densities, are
obtained by the construction of [2]. The Lagrangian
density is a local function of derivatives of the Goldstone field which are
invariant under global $G$-transformations. Upon gauging $G$, these
derivatives are replaced by $G$-covariant derivatives, and the new Lagrangian
density obtained in this way is invariant under $G$-valued gauge
transformations [2]. (An alternative method for obtaining general effective
actions has been advanced in [23], where the nature of locally
gauge invariant effective actions is investigated. This work is also related
to an approach that starts from equivariant cohomology [24].)

It remains to couple to $\G$-valued gauge fields the invariant actions that do
not correspond to invariant Lagrangian densities. As we have seen for the
case of $SU(N),~N\geq3$ in [5], and proven in this paper, these invariant
actions are essentially of the gauged Wess-Zumino-Witten type. It has been
known since the time of their construction [3] that WZW actions cannot, in
general, be made invariant under the complete gauge group $G$. Instead
their gauge variation reproduces the chiral anomaly [13,14]; in
fact, the WZW term was conceived as a generating functional for these chiral
anomalies [3,4,9].

{}From our point of view, it is the invariant actions associated with
non-invariant Lagrangian densities that are of central importance. Following
[4], we wish to produce an action that couples the Goldstone fields to
$G$-valued gauge fields, though we foresee the fact that it will not be
possible to obtain an action invariant under all of the $\G$-valued gauge
transformations. Instead, we shall show that it is always possible to
construct an action whose variation under gauge transformations only involves
the gauge field and not the Goldstone fields.

Remarkably, we also show that invariant actions associated with non-invariant
Lagrangian densities -- appearing for non-trivial de Rham cohomology -- are
equivalently governed by the appearance of anomalies. We show that the fifth de
Rham cohomology of $G/H$ is non-zero only if the group $G$ has anomalies in
at least some of its representations. Furthermore, while the full group $G$
cannot be gauged, the subgroup $H$ always can be completely gauged.

\bigskip

\noindent
{\it Gauging a general element of $H(G/H;R)$.}

\medskip

The forms $\theta ^A$ of (2.2) are minimally coupled to a $\G$-valued gauge
field $A = dx^\mu A_\mu$ and promoted to forms $\bar \theta ^A$ which are
invariant under $G$-valued gauge transformations
acting to the left on $U$.
$$
\bar \theta =
U^{-1} (d +A) U = \bar \theta ^A T^A
\eqno (7.1)
$$
The 1-forms $\bar \theta$ no longer obey the Maurer-Cartan equations; instead,
the connection $\bar \theta$ has curvature, which is related to the curvature
of the $\G$-valued gauge field $A$.
$$
\eqalign{
dA^A + \half f^{ABC} A^B  A^C =  ~F^A \qquad  & \qquad
d \bar \theta ^A + \half f^{ABC} \bar \theta ^B  \bar \theta ^C
=  ~\bar F ^A \cr
\bar F ^A T^A = ~U^{-1} & F ^A T^A U
\cr }
\eqno (7.2)
$$
The covariant derivative with respect to $\bar \theta$ is denoted by
$\bar D$, and $\bar F$ obeys the Bianchi identity $ \bar D \bar F=0 $.
The inner product with $\bar \theta ^A$ is still denoted by $i_A$.
To every differential form $\Omega$, we associate a form $\bar \Omega$ as
follows
$$
\eqalign{
\Omega _{B_1 \cdots B_m} =& {1 \over n!} \Omega _{B_1 \cdots B_m; A_1 \cdots
A_n} \theta ^{A_1} \cdots \theta ^{A_n}\cr
\bar \Omega _{B_1 \cdots B_m} =& {1 \over n!} \Omega _{B_1 \cdots B_m; A_1
\cdots A_n} \bar \theta ^{A_1} \cdots \bar \theta ^{A_n}
\cr}
\eqno (7.3)
$$
There is a general relation between the covariant derivatives of both forms,
given by
$$
\eqalign{
D \Omega _{B_1 \cdots B_m} = & \Sigma _{B_1 \cdots B_m} \cr
\bar D \bar \Omega _{B_1 \cdots B_m} = & \bar \Sigma _{B_1 \cdots B_m}
+ \bar F ^A i_A \bar \Omega _{B_1 \cdots B_m}
\cr}
\eqno (7.4)
$$
Making use of this result, we obtain the gauged form of the hierarchy of
equations of (3.12), which were the basis for the analysis of cohomology. We
start with a closed (globally) $G$-invariant form $\Omega$ of rank 0 and
degree $n$. According to the notation of (7.3), we associate a new form $\bar
\Omega$. This form is no longer closed in general, and we have the following
hierarchy of equations
$$
\eqalign{
d\bar \Omega = \bar F ^A i_A \bar \Omega
\quad \qquad & \cr
\bar D(i_B \bar \Omega ) =\bar F ^A i_A i_B \bar \Omega
\quad \Rightarrow & \quad
i_B \bar \Omega =\bar D \bar \Omega _B - \bar F ^A i_A \bar \Omega _B \cr
\bar D(i_{\{B_2} \bar \Omega _{B_1\}} )=\bar F ^A i_A i_{\{B_1}\bar \Omega
_{B_2\}}
\quad \Rightarrow & \quad
i_{\{B_2} \bar \Omega _{B_1\}} = \bar D \bar \Omega _{B_1 B_2} -\bar F ^A i_A
\bar \Omega _{B_1B_2} \cr
\bar D(i_{\{B_3} \bar \Omega _{B_1B_2\}} )=\bar F^A i_A i_{\{B_1 }\bar \Omega
_{B_2 B_3\}}
 \quad \Rightarrow & \quad
i_{\{B_3} \bar \Omega _{B_1B_2\}} = \bar D \bar \Omega _{B_1B_2B_3}
-\bar F ^A i_A \bar \Omega _{B_1 B_2 B_3} \cr
\cdots &
%\cr
%\bar D(i_{\{B_{m+1}} \bar \Omega _{B_1 \cdots B_m\}} )=\bar F ^A i_A
%i_{\{ B_{m+1}} \Omega _{B_1 \cdots B_m \} }
%\quad \Rightarrow & \quad
%i_{\{B_{m+1}} \bar \Omega _{B_1 \cdots B_m \}} = \bar D \bar \Omega _{B_1
%\cdots B_{m+1}} - \bar F^A  i_A \Omega _{B_1 \cdots B_{m+1}}
\cr}
\eqno (7.5)
$$
Here it is understood that a $\G$-invariant tensor $d_{B_1 \cdots B_m}$ is
added on the right hand side of the second set of equations whenever the
corresponding degree vanishes.

To obtain a $G$-gauge invariant form out of
$\bar \theta$ and $\bar F$, we first construct a
form whose exterior derivative is independent of the Goldstone
field and only depends upon the gauge field. To do so, we introduce a sequence
of forms of rank 0 and degree $n$ as follows $$
\bar \Omega _{(m)}
=  \bar \Omega + \sum _{k=1} ^m (-1)^k  \bar F ^{B_1} \cdots \bar F^{ B_k}
\bar \Omega _{B_1 \cdots B_k}
\eqno (7.6)
$$
The exterior derivatives of these forms are given by
$$
d\bar \Omega _{(m)} = (-1)^m \bar F^{B_1} \cdots \bar F^{B_{m+1}} i_{\{
B_{m+1}} \bar \Omega _{B_1 \cdots B_m\}}
\eqno (7.7)
$$
Of special interest is the case where $\Omega$ and $\bar \Omega$ are
of degree 5, so that the sequence of $\Omega _m$ terminates at $m=2$. The
form $i_{\{ A} \bar \Omega _{BC\}}$ is then of degree 0 and we have shown in
\S 3 that it must be constructed out of the  $d$-symbols of $\G$. In view of
$G$-invariance of the $d$-symbol and the expression for $\bar F^A$ in terms of
$F^A$ given by (7.2), we find that
$$
\eqalign{
d\bar\Omega _{(2)} =& ~d^{ABC} ~\bar F^A \bar F^B \bar F^C  \cr
= & ~d^{ABC} ~F^A F^B F^C  \cr }
\eqno (7.8)
$$
Thus, the right hand side is a function only of the gauge field $A$.

Regarding the form $\bar \Omega _{(2)}$ as a function of the Goldstone field
$U$ and of the gauge field $A$, we can easily obtain a closed form by making
use of the fact that the right hand side of (7.8) no longer depends upon $U$
as in [9,14,22]. $$ \hat \Omega (U,A) =  \bar \Omega _{(2)} (U,A) - \bar \Omega
_{(2)} (1,A)  \qquad \qquad
d\hat \Omega (U,A) =0
\eqno (7.9)
$$
The form $\hat \Omega $ is no longer gauge invariant, and its gauge variation
precisely coincides with the chiral anomaly, which can be seen as follows.
Since the form $\bar \Omega _{(2)} (U,A)$ is gauge invariant, the gauge
variation $\delta \hat \Omega$ of $\hat \Omega$ is given by $\delta \hat
\Omega (U,A) = - \delta \bar \Omega _{(2)} (1,A)$ which may be evaluated using
the explicit expression for $\bar \Omega _{(2)}$ in (7.6) for $U=1$. A
$\G$-valued gauge transformation acts by $\delta A^B = \delta \bar \theta ^B =
- \bar D \epsilon ^B$, and $\delta \bar F^B = \bar D \delta \bar \theta ^B = -
\bar D ^2 \epsilon ^B$, where $\epsilon ^B$ is an infinitesimal $\G$-valued
function of degree 0. With the help of the hierarchy (7.5), we find
$$
\delta \hat \Omega (U,A) = d \biggl \{
\epsilon ^A \bigl ( i_A \bar \Omega - \bar F ^B i_A \bar \Omega _B + \bar F ^B
\bar F^C i_A \bar \Omega _{BC} \bigr ) \biggr \}
\eqno (7.10)
$$
As a result, the gauge variation of the gauged effective action is given by a
four dimensional integral, which coincides with the chiral anomaly : each
term in (7.10) is proportional to some component of $d_{ABC}$. The
definition of $\hat \Omega (U,A)$ is not unique and can be modified by adding
the exterior derivative of any function of $A$ only [14].

\bigskip
\bigskip

\noindent
{\bf Acknowledgements}

I am indebted to Steven Weinberg for stimulating discussions, for critical
questions and for his encouragement. Also, I have
benefited from helpful conversations with Eddie Farhi, Jeffrey Goldstone,
Roman Jackiw,
Roberto Peccei, D.H. Phong, and especially Raja Varadarjan. I thank Scott
Axelrod and S. Wu for drawing my attention to their related work on
equivariant cohomology.

\bigskip
\bigskip

\noindent
{\bf References}

\medskip

\item{[1]} S. Weinberg, Phys. Rev. {\bf 166} (1968) 1568.

\item{[2]} S. Coleman, J. Wess and B. Zumino, Phys. Rev. {\bf 177} (1969)
2239; C.G. Callan, S.~Coleman, J. Wess and B. Zumino, Phys. Rev. {\bf 177}
 (1969) 2247.

\item{[3]} J. Wess and B. Zumino, Phys. Lett.  {\bf 37B} (1971) 95.

\item{[4]} E. Witten,  Nucl. Phys. {\bf B223} (1983) 422.

\item{[5]} E. D'Hoker and S. Weinberg, Phys. Rev. {\bf D50} (1994) 605.

\item{[6]} A. Weil, {\it Collected Works}  (1949);
A. Borel, Ann. Math. (2) {\bf 57} (1953) 115; also in
A. Borel, {\it Collected Papers}, Vol I (Springer Verlag, 1983).

\item{[7]} H. Cartan, in {\it Colloque de Topologie, Centre Belge
de Recherches Math\'e-matiques, Brussels 1950}, (G. Thone, 1950).

\item{[8]} S.S. Chern, {\it Complex
Manifolds without Potential Theory}  (Springer Verlag, 1979).

\item{[9]} L. Bonora and P. Cotta-Ramusino, Phys. Lett. {\bf 107B} (1981) 87;
B. Zumino, in {\it Relativity, groups and Topology II:
Les Houches 1983},
B. De Witt, R. Stora, eds. (North Holland, 1984);
K. Chou, H.Y. Guo, K. Wu and X. Song, Phys. Lett. {\bf 134B} (1984) 67;
J. Manes, R. Stora and B. Zumino, Comm. Math. Phys. {\bf 102} (1985) 157; J.
Manes, Nucl. Phys. {\bf B250} (1985) 369.

\item{[10]} B. De Wit, C.M. Hull and M. Ro\^cek, Phys. Lett. 184B (1987) 233.

\item{[11]} W. Greub, S. Halperin and R. Vanstone, {\it
	Connections, Curvature and Cohomology}, Vol III, (Acad. Press, 1976).

\item{[12]} J. Goldstone and F. Wilczek, Phys. Rev. Lett. {\bf 47}
(1981) 986.

\item{[13]} R. Jackiw, in {\it Current Algebra and Anomalies}, by S.B.
Treiman, R. Jackiw, B. Zumino and E. Witten, Princeton Univ. Press, 1985

\item{[14]} W.A. Bardeen, Phys. Rev. {\bf 184}  (1969) 1848;
H. Georgi and S.L. Glashow, Phys. Rev. {\bf D6} (1972) 429;
D.J. Gross and R. Jackiw, Phys. Rev. {\bf D6}  (1972) 477;
H. Georgi, {\it Lie Algebras in Particle Physics}, (Benjamin/Cummings, 1982).

\item{[15]} Y.-S. Wu, Phys. Lett. {\bf 153B} (1985) 70; C.M. Hull amd B.
Spence, Phys. Lett.
{\bf 232 B} (1989) 204; I. Jack, D.R. Jones, N. Mohammedi and H. Osborn,
Nucl. Phys. {\bf B332} (1990) 359; C.M. Hull and B. Spence, Nucl. Phys. B353
(1991) 379.

\item{[16]} G. 't Hooft, Cargese Summer Institute Lecture Notes, in {\it
Dynamical Gauge Symmetry Breaking}, E. Farhi and R. Jackiw Eds. World
Scientific, Singapore (1982); T. Banks, S. Yankielowicz and A. Schwimmer,
Phys. Lett. {\bf 96B} (1980) 67.

\item{[17]} B.A. Dubrovin, A.T. Fomenko and S.P. Novikov, {\it
	Modern Geometry and Applications}, Vol III (Springer Verlag, 1990);
 M.~Spivak, {\it Differential Geometry, Vol. 5}, Publish or Perish, Inc.
Houston, 1975.

\item{[18]} S.I. Goldberg, {\it Curvature and Homology}, Dover Publications,
Inc., New York, 1982

\item{[19]} E. D'Hoker and J. Goldstone, Phys. Lett. {\bf 158B} (1985) 429.

\item{[20]} F. Wilczek, {\it Fractional Statistics and Anyon
Superconductivity}, World Scientific, Singapore, 1990

\item{[21]} {\it Encyclopedic Dictionary of Mathematics},
                 S. Iyanaga and Y. Kawada, eds. (MIT Press, 1980).

\item{[22]} E. D'Hoker and E. Farhi, Nucl. Phys. {\bf B248} (1984) 59,77.

\item{[23]} H. Leutwyler, ``Foundations of Chiral Perturbation Theory",
Univ. Bern preprint, BUTP-93/24, to appear in Annals of Physics.

\item{[24]} S. Axelrod, Princeton Ph.D. thesis, unpublished (1991);
 S. Wu, J. Geom. Physics, {\bf 10} (1993) 381;
J.M. Figueroa-Farrill and S. Stanciu, {\it Equivariant Cohomology
and Gauged Bosonic Sigma Models}, QMW-PH-94-17, hep-th/9407149 preprint
(1994).

 \end